# Different studies of the global pitch angle of the Milky Way's spiral arms

Jacques P. Vallée

National Research Council Canada, National Science Infrastructure, Herzberg Astrophysics, 5071 West Saanich Road, Victoria, B.C., Canada V9E 2E7

Keywords: Galaxies: spiral - Galaxy: disk - Galaxy: structure - Methods: statistical - ISM: general - ISM: structure

**Abstract.** There are many published values for the pitch angle of individual spiral arms, and their wide distribution (from -3$^{o}$ to -28$^{o}$) begs for various attempts for a single value. Each of four statistical methods used here yields a mean pitch angle in a small range, between -12$^{o}$ and -14$^{o}$ (table 7, Figure 2). The final result of our meta-analysis yields a mean global pitch angle in the MilkyWay's spiral arms of -13.1$^{o}$ ± 0.6$^{o}$.

## 1. Introduction

In a previous study, a persistent large discrepancy was noted in the global pitch angle (Fig.3 in Vallée 2014b), from global spiral arm fits (positional method). Here we query the global pitch angle using three other methods, to see if these means are comparable with the mean from the positional method.

There is a strong support for the concept of a mean global arm pitch angle for a whole spiral galaxy. In some nearby spiral galaxies, pitch angles are measured for each arm, or segment of arm, and statistics on pitch angle have been made. Thus Honig & Reid (2015) used the positions of HII regions to delineate the pitch angles of spiral arms in M51, M74, NGC 1232 and NGC 3184, and they noted that the pitch angle can vary somewhat along segments of an individual arm around a mean value (with arm segments of 5 kpc each, in their Figure 9), and that there is no obvious change of the mean pitch angle with galactocentric distance from the galactic center. Thus the concept of a global spiral arm pitch angle is useful, as an average for a rough galactic model, as an average to compare to other late-type galaxies, etc.

The shape of our Milky Way galaxy is difficult to probe from the position of the Sun in the galactic disk, owing to dust at optical and near-infrared wavelengths (variable absorption of light at different galactic longitudes), and to the need for accurate distances to specific objects (a different distance for each spiral arm). Streaming motions can affect kinematical models, as kinematical distances are not very accurate. Magnetic fields can affect polarization angle and rotation measure; magnetic fields in the Milky Way were recently reviewed by Haverkorn (2015), Heiles & Haverkorn (2012), Vallée (2012), Vallée (2011). An evolution of our knowledge of the interarm distance through the Sun (between the Perseus arm and the Sagittarius arm) was found (Vallée, 2005; Vallée, 2013). An angular offset and a linear separation were found for each chemical tracer, using observed tangents to each spiral arm (Vallée 2014a; Vallée 2014c).

In Section 2, we derive the pitch angle from the development of a *novel* method, using the observed arm tangents to the same arm in two different galactic quadrants ("twin arm method”). In Section 3, we collect, survey and analyse the different published pitch angle values from individual fit to each individual spiral arms (“individual arms method”). In Section 4, we survey and statistically analyze the different pitch angle values from global fits as read from recent publications (“positional method”). In Section 5, we derive the local pitch angle from the interarm separation near the Sun as read in recent publications (“adjacent arms method”). In Section 6, we compare the results from these three other methods with the results from the positional method.

## 2. Twin arm method - pitch angle from the twin tangents to the same arm in different galactic quadrants, for a log-type arm

Here we propose a novel method to extract a common pitch angle for a long arm covering two galactic quadrants. Inner spiral arms can be seen to cross the sun-Galactic Center line, allowing an arm tangent to be measured in Galactic Quadrant IV (longitude $l_{IV} < 360^o$) and another arm tangent in Galactic Quadrant I (longitude $l_I > 0^o$). These twin tangents (for the same long arm) cannot be equally distanced in longitudes from the Sun-Galactic center line, unless the pitch angle is zero. The greater the arm’s pitch angle p, the bigger the difference in angular separation between the twin tangents (unequal longitudes).

This novel method assumes that the arm has a constant pitch angle over the two quadrants. It uses two extremely well-observed galactic longitudes for the tangent to that spiral arm (e.g., Table 1 in Vallée 2014c). It does not use the arm stuff in between (ignoring the intermediate data points in between where the tangents lie).

**Figure 1** shows a triangle composed of the side (a) the line from the arm tangent point to the Galactic Center, the side (b) the arm tangent from the Sun, the side (c) the line from the Sun to Galactic Center. Galactic quadrants I to IV are shown, as well as the Carina-Sagittarius arm, and the Crux-Centaurus-Scutum arm. The angle A between side (b) and side (c) is the angular separation of the arm from the l=0 line, thus it is the galactic longitude (l) in quadrant I and it is ($360^o$ - l) in quadrant IV. The angle C between side (a) and side (b) is ($0.5\,\pi$ + p) in Quadrant I, and it is ($0.5\,\pi$ - p) in Quadrant IV, by construction.

The triangle in Quadrant I for the Sagittarius arm has a side $b_i$ (distance from to the Sun to the spiral arm tangent point, a side $a_I$ (distance from that tangent point to the G.C.), and a side c (distance from G.C. to the Sun), with inside an angle $A_I$ (from the Sun to G.C. side c to the Sun-tangent point side $b_I$ ), an angle $B_I$ (from the Sun to G.C. side c to the G.C. to tangent point side $a_I$ ), and an angle $C_I$ (from side $b_I$ to side $a_I$).

The angle $A_I$ is the galactic longitude $l_I$ , while the angle $\theta_I = 90^o - B_I$ , and the angle $C_I = 90^o$ + pitch angle p. Since the sum of all angles inside a triangle is $180^o$, then it follows that $\theta_I = l_I$ + pitch angle p.

The triangle in Quadrant IV for the Carina arm has a side $b_{IV}$ (distance from to the Sun to the spiral arm tangent point, a side $a_{IV}$ (distance from that tangent point to the G.C.), and a side c (distance from G.C. to the Sun), with inside an angle $A_{IV}$ (from the Sun to G.C. side c to the Sun-tangent point side $b_{IV}$), an angle $B_{IV}$ (from the Sun to G.C. side c to the G.C. to tangent point side $a_{IV}$), and an angle $C_{IV}$ (from side $b_{IV}$ to side $a_{IV}$).

The angle $A_{IV}$ = (360$^o$ - galactic longitude $l_{IV}$), while the angle $\theta_{IV}$ = (90$^o$ + $B_{IV}$), and the angle $C_{IV}$ = (90$^o$ - pitch angle p). Since the sum of all angles inside a triangle is 180$^o$, then it follows that $\theta_{IV}$ = ($l_{IV}$ + pitch angle p - 180 $^o$) .

Similar equations can be made for the dashed triangle in Quadrant I for the Scutum arm, and the dashed triangle in Quadrant IV for the Crux-Centaurus arm.

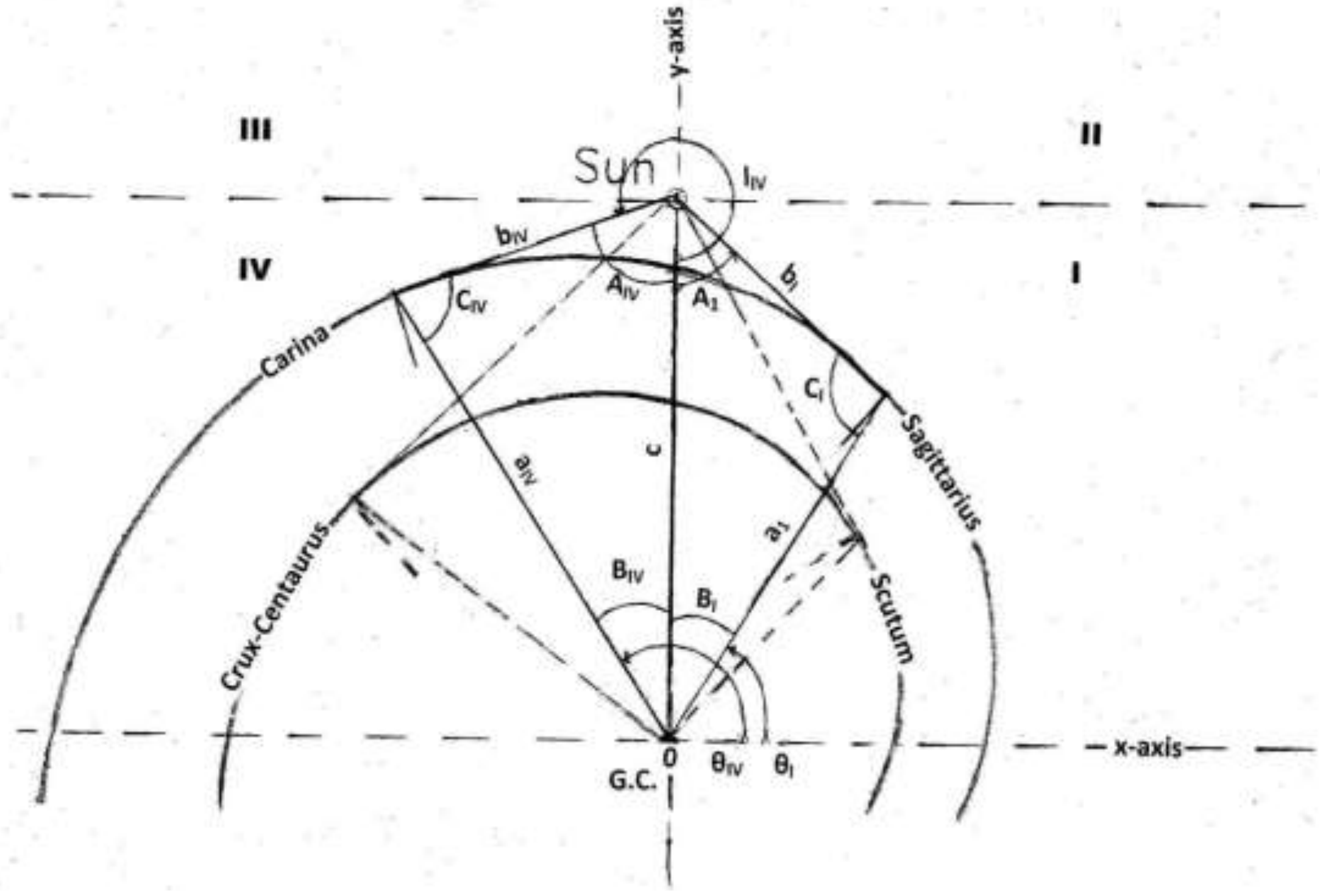

*Figure 1. A sketch of the Sun (circle dot) and the Galactic Center (G.C.). The Solar-based coordinate system has galactic longitudes l and distances b, while the G.C.-based coordinate systems (x-axis, y-axis) has angles $\theta$ and distances a. Four galactic quadrants are shown (I for 0 < l < 90 degrees; II for 90 < l < 180 degrees; III for 180 < l < 270 degrees; IV for 270 <l< 360 degrees).*

Here we will use two laws to yield one equation (Equ. 10) with one unknown and two known variables.

a) From the trigonometric law of sines, a/SinA = c/SinC, thus

$a_I = \sin(l_I)$ . side c / sin (0.5 $\pi$ + p) …in Q I (1)

$a_{IV} = \sin(2\pi - l_{IV})$ . side c / sin (0.5 $\pi$ - p) …in Q IV (2)

and taking the ratio, with sin (0.5 $\pi$ + ε) = sin (0.5 $\pi$ - ε), one gets

$$a_I / a_{IV} = \sin(l_I) / \sin(2\pi - l_{IV}) \quad (3)$$

It can be seen that for a ring (p=0), then the two sides are equal since the two angular separations (longitudes) are equal.

The angle B between side (a) and side (c) is deduced from B =( $\pi$ – A –C), while the complement to the angle B is called $\theta_I$ in Quadrant I. Here $\theta$ is the angle between the galactic x-axis (perpendicular to side c, at the Galactic Center) and the side a (direction to the tangent point), measured counterclockwise from the x-axis. Hence in each quadrant Q:

$$\theta_I = 0.5\pi - B = 0.5\pi - (\pi - A - C) = 0.5\pi - (\pi - l_I - (0.5\pi + p)) = l_I + p \qquad \text{...in Q I (4)}$$

$$\theta_{IV} = 0.5\pi + (-A - C) = 0.5\pi + \pi - (2\pi - l_{IV}) - (0.5\pi - p) = l_{IV} + p - \pi \qquad \text{...in Q IV (5)}$$

and subtracting, one gets:

$$\theta_I - \theta_{IV} = l_I - l_{IV} + \pi \qquad (6)$$

b) From the logarithmic spiral law, one already has (Equation 1 in Vallée 2002):

$$a_I = r_o \exp[\tan(p) . (\theta_I - \theta_o)] \qquad (7)$$

$$a_{IV} = r_o \exp[\tan(p) . (\theta_{IV} - \theta_o)] \qquad (8)$$

and taking the ratio, one gets

$$a_I / a_{IV} = \exp[\tan(p) . (\theta_I - \theta_{IV})] = \exp[\tan(p) . (l_I - l_{IV} + \pi)] \qquad (9)$$

Equating the same ratio $a_I/a_{IV}$, from Equ. 2 and Equ. 4, and taking the log, one gets:

$$\ln[\sin(l_I) / \sin(2\pi - l_{IV})] = \tan(p) . (l_I - l_{IV} + \pi) \qquad (10)$$

This last equation yields p, when the two arm tangents to the same arm (in Quadrant I and Quadrant IV) are known. It is only a function of the observed quantities $l_I$ and $l_{IV}$.

*This novel method employs observed tangents over a wide range in galactic longitudes, and can be used for mid-arm tracers appearing in both quadrants (broad CO, thermal electron, old HII complex, etc).* Numerous, exhaustive galactic longitude data for the arm tangents have been compiled recently in Table 1 of Vallée (2014c).

We will employ them here with Equation (10) above., upon twinning of the Carina arm in Quadrant IV with the Sagittarius arm in Quadrant I.

**Table 1 – Pitch angle from twin arm tangents, for the continuing spiral arm Carina-Sagittarius, with a log-shape arm**

| | Carina spiral arm | | Sagittarius arm | | |
|---|---|---|---|---|---|
| Chemical arm tracer [(a)] | observed tangent in galactic longitude [(a)] (o) | separation in gal. long. (o) | observed tangent In galactic longitude [(a)] (o) | tan(p) | combined pitch p (o) |
| $^{12}CO$ at 8' | 282 | 78 from l=0 | 51 | 0.258 | -14.5 |
| Thermal electron | 283 | 77 from l=0 | 49 | 0.270 | -15.1 |
| Old HII complex | 284 | 76 from l=0 | 51 | 0.240 | -13.5 |
| Dust 240$\mu$ m | 284 | 76 from l=0 | 50 | 0.251 | -14.1 |
| FIR [CII] & [NII] | 287 | 73 from l=0 | 50 | 0.223 | -12.6 |
| median | 284 | 76 from l=0 | 50 | - | -14.1 |
| mean | 284 | 76 from l=0 | 50 | - | -14.0 |
| standard dev.mean | ±0.8 | ±0.8 | ±0.4 | - | ±0.4 |
| | mid-arm tracers ($^{12}CO$, thermal, old HII – top 3 rows) | | | | -14.4 |
| | starforming tracers (dust, FIR - last 2 rows) | | | | -13.4 |

*Note:*
*(a): all galactic longitude data from Table 1 in Vallée (2014c)*

**Table 1** gives the arm tangents to the Carina arm (Quadrant IV) and those for the Sagittarius arm (Quadrant I), and the resulting common pitch angle p, from this last equation. The mean pitch angle is -14.0$^{o}$ here, with an r.m.s. near 0.4$^{o}$.

We again employ Equation (10) above, upon twinning the Crux-Centaurus arm in Quadrant IV with the Scutum arm in Quadrant I.

**Table 2 – Pitch angle from twin arm tangents, for the continuing arm Crux-Centaurus-Scutum, with a log-shape arm**

| | Crux-Centaurus spiral arm | | Scutum spiral arm | | |
|---|---|---|---|---|---|
| Chemical arm tracer [(a)] | observed tangent in galactic longitude [(a)] (o) | separation in gal. long. (o) | observed tangent In galactic longitude [(a)] (o) | tan(p) | combined pitch p (o) |
| $^{12}$CO at 8' | 309 | 51 from l=0 | 33 | 0.212 | -11.9 |
| Thermal electron | 309 | 51 from l=0 | 32 | 0.226 | -12.7 |
| Old HII complex | 309 | 51 from l=0 | 32 | 0.226 | -12.7 |
| FIR [CII] & [NII] | 309 | 51 from l=0 | 30 | 0.255 | -14.3 |
| Sync. 408 MHz | 310 | 50 from l=0 | 32 | 0.215 | -12.2 |
| HI atom | 310 | 50 from l=0 | 29 | 0.259 | -14.5 |
| Dust 240$\mu$ m | 311 | 49 from l=0 | 31 | 0.219 | -12.4 |
| Dust 60$\mu$ m | 311 | 49 from l=0 | 26 | 0.297 | -16.5 |
| Dust 870$\mu$ m | 311 | 49 from l=0 | 31 | 0.219 | -12.4 |
| median | 310 | 50 from l=0 | 31 | - | -12.7 |
| mean | 310 | 50 from l=0 | 31 | - | -13.3 |
| standard dev.mean | ±0.3 | ±0.3 | ±0.7 | - | ±0.5 |
| | mid-arm tracers ($^{12}$CO, thermal, old HII – top rows) | | | | -12.4 |
| | starforming tracers (dust, FIR, etc - bottom rows)[(b)] | | | | -13.2 |

*Notes:*
*(a): all galactic longitude data from Table 1 in Vallée (2014c)*
*(b): excluding the dust 60$\mu$ m tracer (too far away)*

**Table 2** gives the pitch angle results for the Crux-Centaurus arm tangent (Quadrant IV) twinned with the Scutum arm tangent (Quadrant I), separately for each chemical arm tracer. The mean pitch angle is -13.3$^{o}$ here, with an r.m.s. near 0.5$^{o}$.

As observed elsewhere for the Milky Way (Vallée 2014a, Vallée 2014c), starforming tracers (hot dust, say) have a small offset towards the Galactic Center when compared to non-starforming tracers (broad CO, thermal electrons, old HII complex). Being small, this starforming offset would lead to very similar pitch angle results as for a zero offset (within the errors) – see the last rows of these Tables.

## 3. Individual arms method – pitch angle from measurement of individual spiral arms

In the Milky Way galaxy, differing pitch angle values have been published for each individual arm, or segment of arm. Here we carry out a metastudy over 21 recently published papers on individual spiral arms.

**Table 3** shows the literature data, for each individual spiral arm in the Milky Way. For a given arm, the spread of the measured pitch angle is around $8^o$. Most papers employed maser data, which can yield the distance to the masers. A fit to several masers in the same spiral arm can yield its arm pitch angle. Here we use basic or 'blind' statistics, and our analyses over earlier data indicated that our weighted averages and un-weighted averages gave similar results, within their errors. Here we have added a separate line of statistics for different methods, again getting roughly similar results.

Taking all data, statistics can be done here on all the pitch angle results. We find the global median value, done over the median value for each individual arm (columns 1 to 5), to be $-12.1^o$. But if we restrict the statistics to the masers data from the latest results (2013-2015) in Table 3, then we find a global mean pitch of $-12.2^o \pm 1.7^o$.

Each method has its drawbacks. The kinematical method suffers from the need for a precise distance determination. Earlier parallax results have been disputed (Hachisuka et al 2015), notably on the precise atmospheric correction. So far, the parallax method has yielded nothing for two arms (Norma arm, Cygnus +1 arm), these arms being too distant for parallactic detection. In addition, for the parallax method, the Reid et al (2009) paper differs substantially from the Reid (2012) paper, and from the Reid et al (2014) paper, downward for the Perseus arm ($-16.5^o$ vs $-13.0^o$ vs $-9.4^o$) and upward for the Cygnus arm ($-2.3^o$ vs $-12.0^o$ vs $-13.8^o$), possibly due to inclusion or exclusion (gaps) of parts of a spiral arm where there is little massive star formation over a long arc in galactic longitude. Finally, if the parallax measurement is limited to a selected longitude range of a given spiral arm, the value obtained may reflect the pitch angle of that part of the spiral arm, as each portion of the same arm can have a different pitch angle than for that arm's overall value (see Fig. 9 in Honig & Reid 2015).

We can also compare with other nearby spiral galaxies. In a limited study covering 5 selected papers with parallax data published recently (2013-2014), Honig & Reid (2015) found similar pitch angles for Milky Way spiral arms (Scutum, Sagittarius, Perseus, Cygnus) with a mean of $-12.9^o \pm 2.8^o$ (sdm). They also found that the variation of pitch angle among segments of Milky Way spiral arms to be qualitatively similar to those of four other late-type spirals they studied (M51, M74, NGC 1232 and NGC 3184).

**Table 3 – Observed individual pitch angle for each spiral arm in the Milky Way (since 2009).**

| Norma--3kpc arm | Scutum-Crux-Centaurus arm | Carina--Sagitt. arm | Perseus arm | Cygnus arm | Cygnus +1 arm | Method | Data used[a] | Reference |
|---|---|---|---|---|---|---|---|---|
| - | - | - | - | - | -9.3 | kinem | CO data | Fig.3 in Sun et al (2015) |
| - | - | - | - | -14.9 | - | paral | $H_2O$, meth. masers | Fig. 6 in Hachisuka et al (2015) |
| -9.9 | -10.5 | -10.0 | -8.1 | -2.7 | - | kinem | HII and GMC | Tab. 1 in Hou et al (2014) |
| - | -19.8 | - | - | - | - | paral | $H_2O$, meth. masers | Fig. 4 in Sato et al (2014) |
| -6.6 | -13.4 | - | - | - | - | kinem | CO clouds | Tab. 3 in Garcia et al (2014) |
| - | -19.8 | -6.9 | -9.4 | -13.8 | - | paral | $H_2O$, meth. masers | Tab. 2 in Reid et al (2014) |
| - | -11.2 | -9.3 | -14.8 | -11.5 | - | paral | $H_2O$, meth.masers | Tab.2 in Bobylev & Bajkova (2014b) |
| - | - | -7.3 | - | - | - | paral | $H_2O$, meth. masers | Fig. 4 in Wu et al (2014) |
| - | - | - | -9.9 | - | - | paral | $H_2O$ masers | Fig.15 in Choi et al (2014) |
| - | - | -6.2 | - | - | - | paral | $H_2O$ masers | Fig. 6 in Chibueze et al (2014) |
| - | -12.5 | -9.4 | -15.2 | -13.3 | - | paral | $H_2O$, meth. masers | Tab. 1 in Bobylev & Bajkova (2013) |
| - | - | - | -9.5 | - | - | paral | $H_2O$ masers | Fig. 10 in Zhang et al (2013) |
| - | - | - | -17.8 | -11.6 | - | paral | $H_2O$ masers | Fig. 3 in Sakai et al (2012) |
| - | -7.0 | -8.0 | -13.0 | -12.0 | - | paral | $H_2O$, meth. masers | Fig.4 in Reid (2012) |
| - | - | - | - | -12.1 | - | paral | $H_2O$ masers | Fig. 5 in Sanna et al (2012) |
| - | -14.2 | - | - | - | - | kinem | CO clouds | Fig.4 in Dame & Thaddeus (2011) |
| - | - | - | -12.0 | -12.6 | -5.6 | kinem | HII regions | Fig. 7 in Foster & Cooper (2010) |
| -13.5 | -15.6 | -13.6 | -13.5 | - | - | kinem | FIR [CII] and [NII] | Tab. 3 in Steiman-Cameron et al (2010) |
| - | - | -11.2 | - | - | - | paral | $H_2O$ masers | Fig. 6a in Sato et al (2010) |
| -9.2 | -12.5 | -11.1 | -8.4 | - | - | kinem | HII and GMC | Tab. 1 in Hou et al (2009) |
| - | - | - | -16.5 | -2.3 | - | paral | $H_2O$, meth. masers | Fig. 2 in Reid et al (2009) |
| -9.9 | -13.8 | -9.3 | -12.5 | -12.1 | - | -12.1 | Median value for individual arm (all data since 2010) | |
| -10.0 | -13.8 | -9.1 | -12.3 | -11.6 | - | - | Mean value (all data since 2010) | |
| ±2.0 | ±1.4 | ±0.8 | ±1.0 | ±1.2 | - | - | Standard deviation of the mean (all data since 2010) | |
| - | -15.8 | -7.8 | -11.8 | -13.4 | - | -12.2±1.7 | Mean and sdm for parallax method with data 2013-2015[b] | |
| - | -7.0 | -9.6 | -15.4 | -11.9 | - | -11.0±1.8 | Mean and sdm for parallax method with data 2010-2012[b] | |

*Notes:*

*(a): GMC = Giant Molecular Clouds; HII = HII regions; meth. = methanol*

*(b): Drawbacks of the parallax method are discussed in Section 3.*

## 4. Positional method – pitch angle from recent publications on individual tracers

Various independent recent publications have offered a value for the global pitch angle of spiral arms in the Milky Way. Each paper normally covers a small galactic longitude range (typically $90^o$), or a small distance range (typically 5 kpc); a global view ensues when we put all these results together.

Since 1995, we have catalogued the different published results for the Milky Way's arms, allowing us to see some trends with the passage of time, owing to our evolving knowledge. There are not enough recent publications to do robust statistics for each different object group (masers, stars, dust, cosmic rays, HII, bubbles, clouds, CO, etc). Blocks are limited to a minimum of 15 and a maximum of 20 entries, as done earlier in this series of paper. Other papers in this series were: Vallée (1995 – Paper I), Vallée (2002 – Paper II), Vallée (2005 – Paper III), Vallée (2008 – Paper IV), Vallée (2013 – Paper V), Vallée (2014a – Paper VI), Vallée (2014b – Paper VII).

Here we carry out a metastudy over 31 recently published papers on the shape of the Milky Way galaxy. We gathered from published arm data their chosen pitch angle, number of arms used, interarm separation through the sun's location, which tracer was used, the reference to the most telling figure (or table). These data are summarized into blocks. Primary data (pitch angle, interarm at the Sun's position, arm shape and number) are read off from the 31 recently published figures employing a global model (their own model, or one they adapted from others) displaying their chosen arm tracers. Multiple or 'correlated' entries, except the one having appeared in a refereed publication, were deleted. A re-processed entry, taking advantage of new data or a re-calibration, was kept.

**Table 4** shows a block of recently published results for mid-2014. Early published results for 2014 were catalogued in Paper VII.

**Table 5** shows a block of results for late 2014, and early 2015. Where appropriate, some added comments are given in **Appendix A**. Here we do not wish to weight different methods (parallax, kinematical distance, etc). Going over older data would involve re-calibrating the older data reduction method, owing to our better knowledge over time of previously hidden assumptions and decreasing systematic sources of errors – this is outside the scope of this statistical analysis.

A simple statistical analysis of these tables was made. The median values and mean values are computed at the bottom of these tables. The mean value for the pitch angle is $-13.0^o$ (table 4) and $-13.0^o$ (table 5), with a typical r.m.s. near $1^o$.

**Table 4 – Recent studies of the spiral arms in the Milky Way (mid-2014).**

| Pitch Angle (deg.) | No. of arms | Arm shape [(a)] | Inter-arm [(b)] (kpc) | Data used | data read off Figures and references |
|---|---|---|---|---|---|
| -13 | 4 | log | 3.1 | red clumps; red giant stars | Fig.6 in Bobylev et al (2014) |
| -12 | 4 | log | 3.1 [(c)] | $^{13}CO$ associated with masers | Fig.2 in Ren et al (2014) |
| -13 | 4 | log | 3.8 [(c)] | RR Lyrae stars, optical dust | Fig.10 in Schultheis et al (2014) |
| -14 | 4 | log | 3.2 [(c)] | $H_2O$ masers | Fig.14 in Choi et al (2014) |
| -13 | 4 | log | 3.2 [(c)] | methanol masers | Fig.1b in Olmi et al (2014) |
| -12.8 | 4 | log | 3.3 | O & B stars in Centaurus | Fig.7 in Baume et al (2014) |
| -13 | 4 | log | 3.1 | dust extinction near 549nm | Fig.5 in Sale et al (2014) |
| -11; -28 | 6 | log | 5.0 [(c)] | cosmic rays at 1 GeV | Fig.5 in Benyamin et al (2014) |
| -6 | 2 | log | 2.5 [(c)] | open star clusters | Fig.6 in Schmeja et al (2014) |
| -13 | 4 | log | 3.0[(c)] | starforming complexes | Fig.1 in Traficante ey al (2014) |
| -13 | 4 | log | 3,3 | HII regions | Fig.3 in Tremblin et al (2014) |
| -13 | 4 | log | 3.4 | interstellar bubbles | Fig.19 in Beaumont et al (2014) |
| variable | 4 | polyn | 2.7 [(c)] | HII; giant molecular clouds | Fig.11b in Hou & Han (2014) |
| -12.5 | 4 | log | 3.1 | CO reconstruction | Fig.26b in Pettitt et al (2014) |
| -8.2 | 4 | log | 2.7 [(c)] | HII; giant molecular clouds | Fig.10d in Hou & Han (2014) |
| -13 | 4 | log | 3.1 | Median value (all data) | |
| -13.0 | 4 | log | 3.2 | Unweighted mean (all data) | |
| ±1.2 | - | - | ±0.2 | Standard deviation of the mean (all data) | |

*Notes:*

*(a): The arm shape can be logarithmic (log), or polynomial (polyn), or ring or complex.*

*(b): The separation between the Perseus arm and the Sagittarius arm, through the Sun's location.*

*(c): Corrected for 8.0 kpc as the Sun - Galactic Center distance (not the 8.5 value or else as used).*

**Table 5 – Recent studies of the spiral arms in the Milky Way (late-2014 onwards).**

| Pitch Angle (deg.) | No. of arms | Arm shape [(a)] | Inter-arm [(b)] (kpc) | Data used | data read off Figures and references |
|---|---|---|---|---|---|
| -14 | 4 | log | 2.9 [(c)] | methanol and water masers | Fig.3 in Sato et al (2014) |
| -12 | 4 | log | 3.5 [(c)] | massive young stellar objects | Fig.7 in Urquhart et al (2014) |
| -12 | 4 | log | - | $8^{o}$ mid infrared dark cloud | Fig.4 in Goodman et al (2014) |
| -14 | 4 | log | 3.2[(c)] | $H_2O$ masers | Fig.8 in Burns et al (2014) |
| -13 | 4 | log | 3.0 | $H_2O$ masers | Fig.5 in Chibueze et al (2014a) |
| -13 | 4 | log | 3.1[(c)] | [CII] 158 μm FIR emission | Fig.7a in Velusamy & Langer (2014) |
| -13 | 4 | log | 3.1 | OB associations | Fig.4 in Bobylev & Bajkova (2014) |
| -13 | 4 | log | 3.0 | water masers | Fig.5 in Chibueze et al (2014b) |
| -13 | 4 | log | 3.5[(c)] | electron cosmic-rays | Fig.4b in Werner et al (2015) |
| -12 | 1 | log | - | optical HII regions | Fig.4 in Griv et al (2015) |
| -12.5 | 2 | log | 4.8 [(c)] | CO; giant molecular clouds | Fig.5 in Dobbs & Pettitt (2015) |
| -13 | 4 | log | - | young open star clusters | Fig.12 in Carraro et al (2015) |
| -12.8 | 4 | log | 3.2[(c)] | CO; giant molecular clouds | Fig.3 in Sun et al (2015) |
| -13 | 4 | log | 3.4[(c)] | hail from HVC complex C | Fig.1 in Fraternali et al (2015) |
| -13 | 4 | log | - | HI column density | Fig.1 in Duarte-Cabral et al (2015) |
| -14 | 4 | log | 3.1[(c)] | water masers | Fig. 7 in Hachisuka et al (2015) |
| -13 | 4 | log | 3.2 | Median value (all data) | |
| -13.0 | 4 | log | 3.3 | Unweighted mean (all data) | |
| ±0.2 | - | - | ±0.2 | Standard deviation of the mean (all data) | |

*Notes:*

*(a): The arm shape can be logarithmic (log), or polynomial (polyn), or ring or complex.*

*(b): The separation between the Perseus arm and the Sagittarius arm, through the Sun's location.*

*(c): Corrected for 8.0 kpc as the Sun - Galactic Center distance (not the 8.5 value or else as used).*

## 5. Adjacent arms method – pitch angle from the interarm at the Sun's location, for a log-type arm

Two adjacent arms (not a prolongation of the same arm) can be used to infer the joint pitch angle, as shown elsewhere (Paper I). Assuming a logarithmic shape for the arms in the Milky Way, one can derive the link between the interarm separation between two adjacent arms and the logarithmic shape properties (equation 7 in Vallée 1995):

$$2\pi \tan(p) = m \cdot \ln(1 + s) \quad (11)$$

$$\text{with } s = \text{arm separation} / R_{\text{inner arm}} \quad (12)$$

Some simplifying assumptions apply, namely that one uses the same global pitch angle along all arms, that each spiral arm starts at the same galactic radius, and starts at an azimuth that differs by a multiple of $90^o$.

The 'arm separation' is measured along the Sun to Galactic Center line, between the Sagittarius arm ($R_{\text{inner arm}}$ ) and the Perseus arm. The number of spiral arms m =4, while $\pi$ = 3.1416, and the distance of the sun to the Galactic Center is taken as 8.0 kpc.

The location of the Sagittarius arm is read from recently published fits as $R_{\text{inner arm}}$ =7.0 kpc (Fig.3 in Vallée 2014c; Fig. 2 in Vallée 2014a; Fig.1 in Vallée 2014b).

**Table 6** shows recent averaged values of the arm separation, from recent papers in this series. Using the median of column 1, one gets a pitch angle of $-13.3^o$. The use of the mean of column 2 here would give a pitch angle of $-13.2^o$, with a typical r.m.s. near $0.5^o$.

**Table 6. Recent averaged values of the arm separation through the Sun (using a log-shape arm)**

| median | mean | Reference | No. of papers |
|---|---|---|---|
| 3.1 | 3.2 | Table 4 here | 15 |
| 3.2 | 3.3 | Table 5 here | 16 |
| 3.4 | 3.3 | Table 1 of Paper vii | 15 |
| 2.9 | 2.8 | Table 2 of Paper vii | 15 |
| 2.9 | 3.0 | Table 1 of Paper vi | 17 |
| 3.3 | 3.2 | Table 2 of Paper vi | 15 |
| 3.15 | 3.20 | median (all data) | |
| 3.13 | 3.13 | unweighted mean (all data) | |
| ±0.1 | ±0.1 | standard deviation of the mean (all data) | |

## 6. Comparison of different methods to get a global pitch angle

Here we employed four statistical methods: a) the novel ”twin arm method” for the *continuing* arms in two galactic quadrants, assuming a logarithmic arm shape and using 9 individual tracers (Figure 1, tables 1 & 2, Equ. 10); b) the “*individual* arms method” from 21 individual papers (table 3); c) the “positional method” over the Galaxy from 31 individual papers (tables 4 and 5); d) the “*adjacent* arms method”, assuming a logarithmic arm shape (table 6, Equ. 11).

In **Table 7**, we assemble together the best pitch angle values, from each different method. All in all, we find a pitch angle near -13.1°, with an r.m.s. near 0.6°.

If we keep the log-type arm shape (Sections 2 and 5), then the mean global pitch angle is -13.5° with an r.m.s. near 0.4°. If we do not impose a log arm shape (Sections 3 and 4), then the mean global pitch angle is -12.7° with an r.m.s. near 0.5°. If we exclude the highest (-14.0°) and lowest (-12.2°) pitch results (listed in Table 7), then the range of mean pitch angle values narrows considerably, to between -13.0° and -13.3°.

**Table 7. The best pitch angle value, obtained from each different method.**

| pitch angle median | pitch angle mean | s.d.m.[(a)] | method | table / section |
|---|---|---|---|---|
| -14.1° | -14.0° | 0.4° | twin arm - Carina-Sag. arm | Table 1 in Section 2 |
| -12.7° | -13.3° | 0.5° | twin arm - Crux-Scutum arm | Table 2 in Section 2 |
| -12.1° | -12.2° | 1.7° | indiv. arm – recent means | Table 3 in Section 3 |
| -13.0° | -13.0° | 1.2° | positional | Table 4 in Section 4 |
| -13.0° | -13.0° | 1° | positional | Table 5 in Section 4 |
| -13.3° | -13.2° | 0.5° | adjacent arms | Table 6 in Section 5 |
| | -13.1° ±0.6° | | mean of all methods above, with r.m.s. | |
| | -13.5° ±0.4° | | mean (and r.m.s.) for log-shape arm methods (section 2 and section 5) | |
| | -12.7° ±0.5° | | mean (and r.m.s.) for shape-free arm methods (section 3 and section 4) | |

*Note:*
*(a): standard dev. of the mean, not including external systematic errors in the assumptions used*

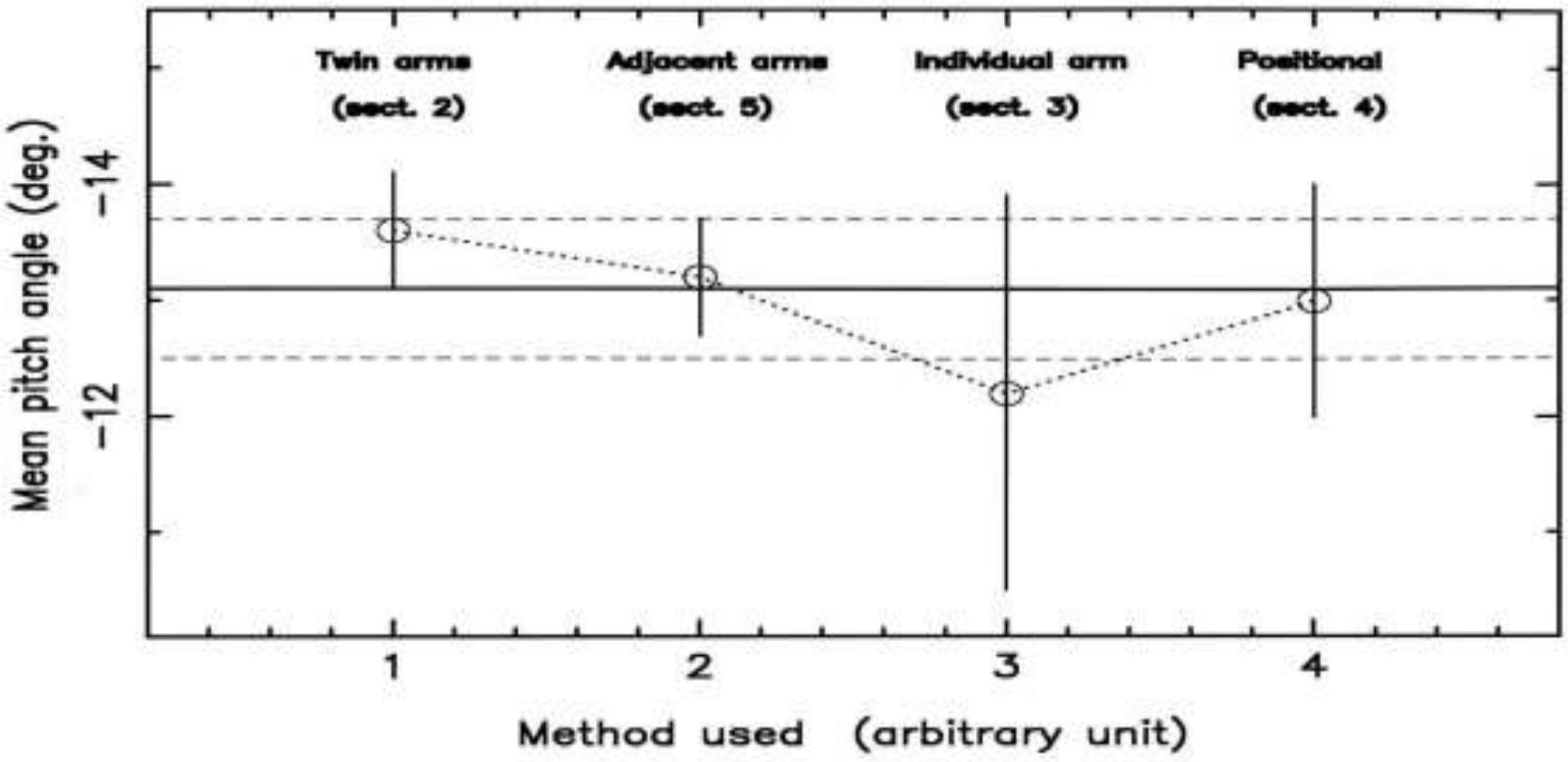

*Figure 2. A comparison of four methods used to get a mean global pitch angle value, using statistics over observational data from the recent literature.*

**Figure 2** shows the range of each method used, and the overall mean (continuous line) and its r.m.s. error (dashed lines) from Table 7. Here one finds a consistency between the four methods, within their respective errors. Hence these 4 different methods statistically agree with each other.

## 7. Conclusion

We have done a recent survey and a meta-analysis, obtained statistical results, employed four methods, and finally gained a consistency on the global pitch angle for the Milky Way. We developed equations to construct a *novel* method to extract the best single pitch angle for a long arm covering two galactic quadrants (Figure 1, Tables 1 and 2). We have analyzed 50 recent papers giving a plethora of published pitch angle values, covering from -3$^o$ to -20$^o$ (Table 3) for individual arms, and covering from -6$^o$ to -28$^o$ (Tables 4 and 5) for the global galaxy. We used the best interarm separation (Table 6) and the log-type arm assumption to get the best mean pitch angle.

Although we find in each method a data set that has a large spread (e.g., -12$^o$ to -15$^o$ in Table 1; -11$^o$ to -16$^o$ in Table 2; -2$^o$ to -19$^o$ in Table 3; -6$^o$ to -28$^o$ in Table 4), the mean and median of each method agrees closely with the mean and median of the other methods.

Thus the mean value of the "twin arms method" (-13.6$^o$ ±0.5$^o$), that of the "adjacent arms method" (-13.2$^o$ ±0.5$^o$) and that of the "individual arms method" (-12.2$^o$ ±1.7$^o$) give an overall mean of -13.0$^o$ ±0.7$^o$ for these 3 methods. This overall mean for these 3 methods can now be agreeably compared with the mean result from recent data for the "positional method" (-13.0$^o$ ±1.0$^o$), answering our query in the beginning of the Introduction.

Using all 4 methods together, our global statistical result (Table 7) is a global pitch angle near -13.1$^o$ ± 0.6$^o$. It is now hard to see how the global pitch angle in the Milky Way could be claimed to be outside the range from -12.0$^o$ to -14.0$^o$ (Figure 2). Any such claim should be scrutinized thoroughly, notably for hidden assumptions, or incomplete calibrations, or strong weighting of some observations (for example, imposing 2 major arms, or choosing a simplistic dust removal, see Vallée 2014b).

## Acknowledgements.

The figure production made use of the PGPLOT software at the NRC Canada in Victoria.

## Appendix A.

Benyamin et al (2014) used a combination of a 2-arm system with a $11^{o}$ pitch angle mixed with a 4-arm system with a $28^{o}$ pitch angle. In their Fig. 5, there is no Scutum arm near longitude $30^{o}$, and no Perseus origin arm near longitude $338^{o}$. Their predicted longitudes for the Carina ($288^{o}$) and Crux ($314^{o}$) arms are outside the observed longitudes using other arm tracers ($284^{o}$ and $310^{o}$).

Dobbs & Pettitt (2015) added to their 2-arm model some long branches (the "Local Arm" and the "Inner Ridge" in their Fig. 5a), both being long (turning more than $360^{o}$ around the Galactic Center). Their "Local Arm" does not include the Sun (it being about 2 kpc away in the Galactic Anticenter direction). Without these 2 long branches, they could not fit the CO longitude-velocity diagram (their Fig.5b).

Griv et al (2015, their fig.4) employed HII regions near the Sun to fit a 1-arm structure. The 1-arm structure does not fit the known Perseus arm location (about 2.5 kpc away from the Sun), nor the known Sagittarius arm location (about 0.5 kpc away from the Sun).

Hou and Han (2014) tried logarithmic and polynomial arm shapes, as well as fixed and variable pitch angles (their fig.12), and fitted 3-arm, 4-arm, and 5-arm models. Their 'far 3-kpc arm" near $l=338^{o}$ is actually the 'Perseus-origin" arm.

Pettitt et al (2014) found that models with 2 arms cannot reproduce all the observed features, and that the 4-arm models require a short local armlet near the Sun (their fig. 26b).

Schmeja et al (2014) used a 2-arm model, which has many turns around the Galactic Center before reaching the Sun; they did not try to fit a 4-arm model. Their model has twice as many predicted arm tangents as seen from the Sun; most of these extra arm tangents are not seen observationally.

Sun et al (2015) used kinematic CO data to get indirect distances, with a scaling to position the High Mass Star Forming Regions in the arms (their Section 3), and got a pitch angle of $-9.3^{o}$. Foster & Cooper (2010) previously used the locations of HII regions in the same longitude range (100 to 150 degrees) at distances from 10 to 16 kpc and fitted a very similar arm there (in green in their Fig.7) with a smaller pitch angle value.

## References:

Baume, G., Rodriguez, M.J., Corti, M.A., Carraro, G., Panei, J.A., " A deep and wide-field view of the IC 2944 / 2948 complex in Centaurus", 2014, MNRAS, v443, p411-422.

Beaumont, C.N., Goodman, A.A., Kendrew, S., Williams, J.P., Simpson, R., "The Milky Way project: leveraging citizen science and machine learning to detect interstellar bubbles", 2014, ApJ Suppl Ser., v214, a3, p1-18.

Benyamin, D., Nakar, E., Piran, T., Shaviv, N., " Recovering the observed B/C ratio in a dynamic spiral-armed cosmic ray model", 2014, ApJ., v782, a34, p1-11.

Bobylev, V.V., Bajkova, A.T., " Estimation of the pitch angle of the galactic spiral pattern", 2013, Astron Letters, v39, no.11, p759-764.

Bobylev, V.V., Bajkova, A.T., “The Milky Way spiral structure parameters from data on masers and selected open clusters”, 2014, MNRAS, 437, 1549-1553.

Bobylev, V.V., Bajkova, A.T., 2014, “The Gould belt, the deVaucouleurs-Dolidze Velt, and the Orion Arm”, 2014, Astron. Letters, v40, p783-792.

Bobylev, V.V., Mosenkov, A.V., Bajkova, A.T., Gontcharov, G.A., “Investigation of the galactic bar based on photometry and stellar proper motions:, 2014, Astron. Lett., v40, p86-90.

Burns, R.A., Nagayama,T., Handa, T., Omodaka, T., Nakagawa, A., Nakanishi, H., Hayashi, M., Shizugami, M., “Trigonometric distance and proper motion of IRAS 20056+3350: a massive starforming region on the solar circle”, 2014, ApJ, v797, a39, p1-9.

Carraro, G., Vazquez, R.A., Costa, E., Ahumada, J.A., Giorgi, E.E., “The thickening of the thin disk in the third galactic quadrant”, 2015, Astron J., v149, a12, p1-17

Chibueze, J.O., Omadaka, T., Handa, T., Imai, H., Kurayama, T., Nagayama, T., Sunada, K., Nakano, M., Hirota, T., Honma, M., “ Astrometry and spatio-kinematics of $H_2O$ masers in the massive star forming region NGC 6334I(North) with VERA”, 2014a, ApJ, v784, a114, p1-8.

Chibueze, J.O., Sakanoue, H., Nagayama, T., Omodaka, T., Handa, T., Kamezaki, T., et al, “Trigonometric parallaxes of IRAS 22555+6213 with VERA: 3-dimensional view of sources along the same line of sight”, 2014b, Publ Astr Soc Japan, v66, p104-113 .

Choi, Y.K., Hachisuka, K., Reid, M.J., Xu, Y., Brunthaler, A., Menten, K.M., Dame, T.M., Trigonometric parallaxes of star forming regions in the Perseus spiral arm”, 2014, ApJ, v790, a99, p1-16.

Dame, T.M., Thaddeus, P., “A molecular spiral arm in the far outer Galaxy”, 2011, ApJ, v734, L24, p1-4.

Dobbs, C., Pettitt, A., “Cloud and star formation in spiral arms”, 2015, in “Lessons from the Local Group”, ed. by Freeman, K., Elmegreen, B., Block, D., Woolway, M., Dordrecht: Springer, p. 147-156.

Duarte-Cabral, A., Acreman, D.M., Dobbs, C.L., Mottram, J.C., Gibson, S.J., Brunt, C.M., Douglas, K.A., “Synthetic CO, $H_2$ and HI surveys of the Galactic 2$^{nd}$ Quadrant, and the properties of molecular gas”, 2015, MNRAS, v447, p2144-2158

Foster, T., Cooper, B., “Structure and dynamics of the Milky Way: the evolving picture”, 2010, Astron Soc Pac Conf Ser., v438, p16-30.

Fraternali, F., Marasco, A., Armillota, L., Marinacci, F. “Galactic hail: the origin of the high-velocity cloud complex C”, 2015, Monthly Not Roy Astr Soc., v447, L70-L74.

Garcia, P., Bronfman, L., Nyman, L.-A., Dame, T.M., Luna, A., “Giant molecular clouds and massive star formation in the southern Milky Way”, 2014, ApJ Suppl Ser., v212, A2, p.1-33

Goodman, A.A., Alves, J., Beaumont, C.N., Benjamin, R.A., Borkin, M.A., Burkert, A., Dame, T.M., Jackson, J., Kauffmann, J., Robitaille, T., Smith, R.J., “The bones of the Milky Way”, 2014, ApJ, v797, A53, p1-13.

Griv, E., Jiang, I.-G., Russeil, D., “Parameters of the galactic density-wave spiral structure: line-of-sight velocities of 156 star-forming regions”, 2015, New Astronomy, v35, p40-47.

Hachisuka, K., Choi, Y.K., Reid, M.J., Brunthaler, A., Menten, K.M., Sanna, A., Dame, T.M.,”Parallaxes of star forming regions in the outer spiral arms of the Milky Way”, 2015, ApJ, v800, a2, p1-8.

Haverkorn, M., “Magnetic fields in the Milky Way”, 2015, Astrophys. Space Sci. Lib., v407, p483-506.

Heiles, C., Haverkorn, M., "Magnetic fields in the multiphase interstellar medium", 2012, Space Sci Rev, v166, p293-305.

Honig, Z.N., Reid, M.J., "Characteristics of spiral arms in late-type galaxies", 2015, ApJ, v800, a53, p1-8.

Hou,L.G., Han, J.L., "The observed spiral structure of the Milky Way", 2014, A&A, v569, a125, p1-23.

Hou,L.G., Han, J.L., Shi, W.B., "The spiral structure of our Milky Way galaxy", 2009, A&A, v499, p473-482.

Olmi, L., Araya, E.D., Hofner, P., Molinaru, S., Morales Ortiz, J., Moscadelli, L., Pestalozzi, M., "Discovery of weak 6.7-GHz $CH_3OH$ masers in a sample of high-mass Hi-GAL sources", 2014, A&A, v566, a18, p1-16.

Pettitt, A.R., Dobbs, C.L., Acreman, D.M., Price, D.J., "The morphology of the Milky Way – I. Reconstructing CO maps from simulations in fixed potentials", 2014, MNRAS, v444, p919-941.

Reid ,M.J., "Galactic structure from trigonometric parallaxes of star forming regions", 2012, Proc. IAU Symp., v289, p.188-193.

Reid, M.J., Menten, K.M., Brunthaler, A., Zheng, X.W., Dame, T.M., Xu, Y., Wu, Y., Zhang, B., Sanna, A., Sato, M., Hachisuka, K., Choi, Y.K., Immer, K., Moscadelli, L., Rygl, K.L., Bartkiewicz, A., " Trigonometric parallaxes of high mass star forming regions: the structure and kinematics of the Milky Way", 2014, ApJ, v783, a130, p1-14.

Reid, M.J., Menten, K.M., Zheng, X.W., Brunthaler, A., Moscadelli, L., Xu, Y., et al, "Trigonometric parallaxes of massive star forming regions. VI. Galactic structure, fundamental parameters, and noncircular motions", 2009, ApJ, v700, p137-148.

Ren, Z., Wu, Y., Liu, L., Li, L., Li, D., Ju, B., " A CO observation of the galactic methanol masers", 2014, A&A, v567, a40, p1-31.

Sakai, N., Honma, H., Sakanoue, H., Kurayama, T., Shibata, K.M., Shizugami, M., "Outer rotation curve of the Galaxy with VERA I: trigonometric parallax of IRAS 05168+3634", 2012, Publ. Astr. Soc. Japan, v64, a108, p1-11.

Sale, S.E., Drew, J.E., Barentsen, G., Farnhill, H.J., Raddi, R., Barlow, M.J., Eisloffel, J., Vink, J.S., Rodriguez-Gil, P., Wright, N.J., "A 3D extinction map of the Northern Galactic Plane based on IPHAS photometry", 2014, MNRAS, v443, p2907-2922.

Sanna, A., Reid, M.J., Dame, T.M., Menten, K.M., Brunthaler, A., Moscadelli, L., Zheng, X.W., Xu, Y., "Trigonometric parallaxes of massive star forming regions. IX. The Outer arm in the first quadrant", 2012, ApJ, v745, a82, p1-7.

Sato, M., Hirota, T., Reid, M.J., Honma, M., Kobayashi, H., Iwadate, K., Miyaji, T., Shibaya, K.M., "Distance to G14.33-0.64 in the Sagittarius spiral arm: $H_2O$ maser trigonometric parallax with VERA", 2010, Publ. Astron. Soc. Japan, v62, p287-299.

Sato, M., Wu, Y.W., Immer, K., Zhang, B., Sanna, A., Reid, M.J., Dame, T.M., Brunthaler, A., Menten, K.M., "Trigonometric paralaxes of star forming regions in the Scutum spiral arm", 2014, ApJ, v793, a72, p1-15.

Schmeja, S., Kharchenko, N.V., Piskunov, A.E., Roser, A.E., Schilbach, E., Froebrich, D., Scholz, R.-D., "Global surveys of star clusters in the Milky Way – III. 139 new open clusters at high galactic latitudes", 2014, A&A, v568, a51, p1-9.

Schultheis, M., Chen, B.Q., Jiang, B.W., Gonzalez, O.A., Enokiya, R., Fukui, Y., Torii, K., Rejkuba, M., Minnitti, D., "Mapping the Milky Way bulge at high-resolution: the 3D dust extinction, CO, and X factor maps", 2014, A&A, v566, a120, p1-11.

Steiman-Cameron, T.Y., Wolfire, M., Hollenbach, D., "COBE and the galactic interstellar medium: geometry of the spiral arms from FIR cooling lines", 2010, ApJ, v722, p1460-1473.

Sun, Y., Xu, Y., Yang, J., Li, F.C., Du, F.C., Zhang, S.B., Zhou, X., "A possible extension o the Scutum-Centaurus arm into the outer second quadrant", 2015, ApJ Lett., v798, L27, p1-5.

Traficante,A., Paladini, R., Compiegne, M., Alves, M.I., Cambresy, L., Gibson, S.J., et al, "The pros and cons of the inversion method approach to derive D3D dust emission properties in the ISM: the HI-GAL field centred on (l,b)=($30^o$, $0^o$)", 2014, MNRAS, v440, p3588-3612.

Tremblin, P., Anderson, L.D., Didelon, P., Raga, A.C., Minier, V., Ntormousi, E., Pettitt, A., Pinto, C., Samal, M., Schneider, N., Zavagno, Z., "Age, size, position of HII regions in the Galaxy", 2014, A&A, v568, a4, p1-7.

Urquhart, J.S., Moore, T.M., Csengeri, T., Wyrowski, F., Schuller, F., Hoare, M.G., et al, "ATLASGAL – towards a complete sample of massive star forming clumps", 2014, MNRAS, v443, p1555-1586.

Vallée, J.P., " The Milky Way's Spiral Arms Traced by Magnetic Fields, Dust, Gas, and Stars", 1995, ApJ, v.454, pp.119-124 (Paper I).

Vallée, J.P., « Metastudy of the Spiral Structure of Our Home Galaxy", 2002, ApJ, v.566, pp. 261-266 (Paper II).

Vallée, J.P., « The Spiral Arms and Interarm Separation of the Milky Way: An Updated Statistical Study", 2005, AJ, v.130, pp. 569-575 (Paper III).

Vallée, J.P., « New Velocimetry and Revised Cartography of the Spiral Arms in the Milky Way—A Consistent Symbiosis", 2008, AJ, v.135, pp. 1301-1310 (Paper IV).

Vallée, J.P., "Magnetic field in the Galactic Universe, as observed in supershells, galaxies, intergalactic and cosmic realms", 2011, New Astron Rev., 55, 91-154.

Vallée, J.P., "Magnetic Milky Way", 2012, Europ. Astron. Soc. Publ. Ser., v56, p81-86.

Vallée, J.P., "A synthesis of fundamental parameters of spiral arms, based on recent observations in the Milky Way", 2013, IJAA, v.3, pp.20-28 (Paper V).

Vallée, J.P., "The spiral arms of the Milky Way: the relative location of each different arm tracer, within a typical spiral arm width", 2014a, Astron,J., v.148, A5, p.1-9 (Paper VI).

Vallée, J.P., "On a persistent large discrepancy in some observed parameters of the spiral arm in the Milky Way – a statistical and modelling analysis", 2014b, MNRAS, v442, p2993-2998 (Paper VII).

Vallée, J.P., "Catalog of observed tangents to the spiral arms in the Milky Way galaxy", 2014c, ApJ Suppl Ser., v215, a1, p1-9

Velusamy, T., Langer, W.D., "Origin and z-distribution of galactic diffuse [CII] emission", 2014, Astron. & Astrophys., v572, a45, p1-20.

Werner, M., Kissmann, R., Strong, A.W., Reimer, O., "Spiral arms as cosmic ray source distributions", 2015, Astroparticle Physics., v64, p18-33.

Wu, Y.W., Sato, M., Reid, M.J., Moscadelli, L., Zhang, B., Xu, Y., Brunthaler, A., Menen, K.M., Dame, T.M., Zheng, X.W., "Trigonometric parallaxes of star form ing regions in the Sagittarius spiral arm", 2014, A&A, v566, a17, p1-26.

Zhang, B., Reid, M.J., Menten, K.M., Zheng, X.W., Brunthaler, A., Dame, T.M., Xu, Y., "Parallaxes for W49N and G048.60+0.02: distant star forming regions in the Perseus spiral arm", 2013, ApJ, v775, a79, p1-13.